\documentclass[a4paper,showpacs,amsmath,amssymb,12pt]{article}
\usepackage{txfonts}
\usepackage{mathrsfs}
\usepackage{amsfonts}
\usepackage{indentfirst}
\usepackage{amssymb}
\usepackage{latexsym}
\usepackage{graphicx}
\usepackage[english]{babel}
\input epsf.sty

\title{Note on Generalized Janus Configurations}
\author{Bin Chen and Zhi-bo Xu\footnote{Email:bchen01,xuzhibo@pku.edu.cn}\\
{\small Department of Physics} \\
{\small and State Key Laboratory of Nuclear Physics and Technology,}\\
{\small Peking University, Beijing 100871, P.R.China}\\\\
Chang-yong Liu\footnote{Email:lcy@itp.ac.cn}\\
{\small Institute of Theoretical Physics}\\
{\small Chinese Academy of Science, Beijing 100080, P.R. China} \\
{\small and Graduate University of Chinese Academy of Science,}\\
{\small Beijing 100080, P.R. China}}
\date{}

\begin{document}
\maketitle
\begin{abstract}
We study several aspects of generalized Janus configuration, which
includes a theta term. We investigate the vacuum structure of the
theory and find that unlike the Janus configuration without theta
term there is no nontrivial vacuum. We also discuss BPS soliton
configuration both by supersymmetry analysis and from energy
functional. The half BPS configurations could be realized by
introducing transverse (p,q)-strings in original brane
configuration corresponding to generalized Janus configuration. It
turns out the BPS soliton could be taken as modified dyon. We
discuss the solution of half BPS equations for the sharp interface
case. Moreover we construct less supersymmetric Janus
configuration with theta term.
\end{abstract}

\newpage
\section{Introduction}
   %After the Ads/CFT correspondence discovered by
   %Maldacena \cite{Mal}, many generalizations of the correspondence have been

 Janus configuration is a kind of field theory with spatially
 dependent coupling constant. Its discovery was  motivated by the
 attempt of generalizing AdS/CFT correspondence\cite{Mal} to the case with spatially varying dilaton.
   The original Janus supergravity solutions is an one-parameter family solutions
   of dilatonic deformation of $AdS_5$, which breaks all supersymmetry
   but    preserves full R-symmetry\cite{BGH}. Later on , the other cases
   preserving
   part of R-symmetry and supersymmetry were discovered \cite{CFKS,CK,HEG0612}.
   Especially, the half BPS Janus solutions with global symmetry $OSP(4|4)$
   were discovered\cite{GR,HEG0724,HEG0722}. The dual gauge
   theories of the Janus solutions are supersymmetric deformations of $N=4$ super
   Yang-Mills theory with coupling constant depending on one spatial
   coordinate \cite{CFKS,HEG0613}\footnote{For a field theory with coupling constant depending
   on a lightlike coordinates and its AdS/CFT correspondence, see \cite{Chu}.}. Some properties of half
   supersymmeric Janus Yang-Mills theory including vacuum structure
   and BPS configuration were discussed in \cite{KKL}.
   For the vacuum structure, it was found that except for the ordinary Coulomb phase where the
   scalars are homogenous and diagonal, there are additional vacuum
   structure characterized by Nahm equations. The BPS configurations
   of supersymmetric Janus field theory include 1/2 BPS magnetic monopole and 1/4 BPS dyonic
   monopole.

   However all of the above Janus field theory do not
   contain a theta term. This sounds strange since one can obtain Janus solution in Type IIB
   supergravity with both dilaton and  axion through an $SL(2,R)$
   transformation. Recently, this puzzle was solved. In \cite{GW1}, the authors obtained generalized Janus
   configuration with spatially varying theta
   angle. The global symmetry is still
   $OSP(4|4)$ but its embedding in $PSU(4|4)$ is inequivalent
   to the one in Janus configuration without $\theta$-term. In order to make the $\theta$
   depend on one spatial coordinate, the embedding  must also depend
   on the spatial coordinate. On the other hand, Janus configuration is closely related to
   the field theory with boundary\cite{GW2,GW3}. This could be seen from the realization
    of half-supersymmetric Janus by brane configuration.   The corresponding
   brane configuration is $N$ D3-branes ending on $k$ successive five branes.
   The limit that $k$ become large and the five-branes are closely spaced corresponds
    to the generalized Janus configuration with
   arbitrary $y$-dependent  $\psi$, which is related to the Yang-Mills coupling
   $\tau=\frac{\theta}{2\pi}+\frac{4\pi i}{e^2}=a+4\pi De^{2i\psi}$ with
   $a$ and $ D$  being constants. Actually in the brane realization, the $a$ and $D$
   could be determined by the brane configurations\cite{GW1}.
   For example,  for the configuration of D3's ending on NS5 branes one has $a=-4\pi D$. However there
   are no such constraints in Janus field theory. In this sense, Janus field
   theory seems to be more general than the brane configuration.
   However such brane configuration may help us to understand
   physics of Janus configuration.

   In this paper, we try to study several aspects of the Janus configurations with theta term.
   We first study its moduli space and BPS configurations. We show that unlike the Janus configuration
   without theta term there is no nontrivial vacuum. We obtain the BPS soliton equations by
   imposing more projective conditions on the supersymmetry parameter in the field theory. We figure out
   the brane  picture corresponding to  half BPS soliton, which requires the introduction of (p,q)-string
   extending along one of transverse directions of D3-branes.
   It turns out that the half BPS soliton could be taken as modified dyon and 1/4-BPS soliton
   could be taken as modified string junction. We discuss the half-BPS solution in the sharp
   interface case. Finally
   we turn to construct generalized Janus configuration with less
   supersymmetries and find that there is no nontrivial vacuum.

     The organization of this paper is as follows. In section 2, we
   give a brief review  of Janus field theory including
   a spatial varying theta term, and then discuss its vacuum structure. This will help us to set up our convention.
   In section 3, we study the  BPS solutions of generalized Janus
   configuration and their corresponding
   brane configurations.
   In section 4, we construct the less supersymmetric Janus field
   theory. We end with some conclusions and discussions.

\section{Janus Configuration With Theta-angle and its Vacuum}

   The four dimensional $N=4$ supersymmetric Yang-Mills
   theory allows a deformation which result in a field theory with a spatially-dependent
   coupling constant and  half supersymmetries\cite{CFKS,HEG0613}. The deformed field theory is called Janus configuration.
   The Janus configuration has
   been extended to include a spatially varying $\theta$
   angle\cite{GW1}. We start from a brief review of this so-called generalized Janus configuration.
   The unperturbed $N=4$ supersymmetric Yang-Mills theory
   could be written in ten-dimensional notation
\begin{equation}
   I=\int d^4x\frac{1}{e^2}\textrm{Tr}\left(\frac{1}{2}F_{IJ}F^{IJ}-i\overline{\Psi}
   \Gamma^ID_I\Psi\right)
\end{equation}
  and the supersymmetry transformations are
\begin{eqnarray}
  \delta_0 A_I&=&i\overline{\epsilon}\Gamma_I\Psi\\
  \delta_0\Psi&=&-\frac{1}{2}\Gamma^{IJ}F_{IJ}\epsilon
\end{eqnarray}
  where $I,J=0,1,2,\cdots,9,\ A_I=A_I^aT_a,
  \textrm{Tr}(T_aT_b)=-\delta_{ab}/2,\ T_a^\dagger =-T_a $.
  The gamma matrices $\Gamma^I$'s are in Majorana
  representation, where $\{\Gamma^I,\Gamma^J\}=2g^{IJ}$ with
  signature $(-++...+)$. The gaugino field $\Psi$ and the
  supersymmetry transformation parameter $\epsilon$ are Majorana-Weyl
  spinors, obeying
  $\overline{\Gamma}\epsilon=\epsilon,\overline{\Gamma}\Psi=\Psi$,
where $\overline{\Gamma}=\Gamma_{012...9}$.
  All the fields are defined in  four-dimensional spacetime
  $x^0,x^1,x^2,x^3$.  Without losing generality, we pick up
  one spacial coordinate $y=x^3$ on which the coupling constant and $\theta$-angle
  depend.  The components of 10-dimensional gauge fields
  $A_I$ with $I=4,5...9$ correspond to six four-dimensional scalar fields $X_I$.
  In this case, the notations $F_{IJ},F_{\mu I}$ mean that
  $F_{IJ}=[X_I,X_J], F_{\mu I}=D_\mu X_I$ when $I,J=4,5...9$. For simplicity, we use
  the ten-dimensional notations $F_{IJ}$ in the above sense.
  The theta term is
\begin{eqnarray}
  I_{\theta}&=&-\frac{1}{32\pi ^2}\int d^4x \theta(y)\varepsilon^{\mu\nu\alpha\beta}
  \textrm{Tr}F_{\mu\nu}F_{\alpha\beta}\nonumber\\
  &=&\frac{1}{8\pi^2} \int d^3xdy
  \frac{d\theta}{dy}\varepsilon^{\mu\nu\lambda} \textrm{Tr}\left(A_{\mu}
  \partial_{\nu}A_{\lambda}+\frac{2}{3}A_{\mu}A_{\nu}A_{\lambda}\right).
\end{eqnarray}
where $\varepsilon^{0123}=-1, \varepsilon^{012}=1$

  The original Janus configurations without theta angle which have
  eight supercharges
  is given by modifying the action and supersymmetry
  transformation such that  half of
  supersymmetries are preserved. In this case,
  the supersymmetry transformation parameter $\epsilon$ does not depend on $y$.
  However, $\epsilon$ should depend on $y$ in order to add the
  theta term. Since the supersymmetry transformation is global,
  any two unbroken supersymmetry transformations should close into a translation
  along $x^0, x^1, x^2$. This requires
\begin{eqnarray}
  \frac{d}{dy}\overline{\epsilon}\Gamma^{\mu}\tilde\epsilon=0,\ \mu=0,1,2,\ \
  \overline{\epsilon}\Gamma^3\tilde\epsilon=0.\label{susy}
\end{eqnarray}
  There are also other global symmetries. The original symmetry $SO(1,9)$
  is broken to $W=SO(1,2)\times SO(3)_X\times SO(3)_Y$. The $SO(1,2)$ is the
  Lorentz symmetry of spacetime in $x^0, x^1, x^2$. The $SO(3)_X$ acts on
  $X_4,X_5,X_6$ while the $SO(3)_Y$ acts on
  $X_7,X_8,X_9$. Taking both the global symmetries and dimension analysis into account,
  we can make the
  following ansatz for the modified  supersymmetry and action
  transformation.
\begin{eqnarray}
  \delta_1\Psi=
   \frac{-1}{2}\left(\Gamma^a X_a(s_1\Gamma_{456}+s_2\Gamma_{789})
   +\Gamma^p
   X_p(t_1\Gamma_{456}+t_2\Gamma_{789})\right)\epsilon
\end{eqnarray}
\begin{eqnarray}
   I^{\prime}&=& \int d^4x\frac{i}{e^2}\textrm{Tr}\overline{\Psi}\left(\alpha\Gamma_{012}
   +\beta\Gamma_{456}+\gamma\Gamma_{789}\right)\Psi\\
   I^{\prime\prime}&=&\int d^4x\frac{1}{e^2}\left(u\varepsilon^{\mu\nu\lambda}
   \textrm{Tr}(A_{\mu}\partial_{\nu}A_{\lambda}
   +\frac{2}{3}A_{\mu}A_{\nu}A_{\lambda})\right.\nonumber\\
   &&\left.+\frac{v}{3}\varepsilon^{abc}\textrm{Tr}X_{a}[X_{b},X_{c}]
   +\frac{w}{3}\varepsilon^{pqr}\textrm{Tr}X_{p}[X_{q},X_{r}]\right)\\
   I^{\prime\prime\prime}&=&\int d^4x\textrm{Tr}\left(\frac{r}{2e^2}X_aX^a
   +\frac{\tilde r}{2e^2}X_pX^p\right)
\end{eqnarray}
where $\mu, \nu, \lambda=0, 1, 2, a, b, c=4, 5, 6, p, q, r=7, 8,
9$ and $\varepsilon$ are antisymmetric tensors normalized to
$\varepsilon^{012}=\varepsilon^{456}=\varepsilon^{789}=1$. All
parameters depend on $y$. Using the condition that the Lagrangian
preserves half of supersymmetries, we can obtain the following
equations:
\begin{eqnarray}
  &&\frac{d\epsilon}{dy}=\alpha\Gamma_{0123}\epsilon\label{epsiequation}\\
  &&\overline{\epsilon}((s_1+2\beta)B_1+(s_2-2\gamma)B_2+q)=0\label{solution1}\\
  &&\overline{\epsilon}((t_1-2\beta)B_1+(t_2+2\gamma)B_2+q)=0\\
  &&\overline{\epsilon}(4\alpha B_0+2\beta B_1+2\gamma B_2-q)=0\\
  &&u=-4\alpha,\ \ \ v=-4\beta,\ \ \ w=-4\gamma\label{solution3}\\
  &&\overline{\epsilon}((-2\beta'-4\gamma\alpha)B_1
  +(2\gamma'-4\beta\alpha)B_2)=\overline{\epsilon}\lambda\label{solution2}\\
 && r=\lambda+2\beta^2+2\gamma^2-\frac{q^2}{2}-q'\\
  &&\tilde r=-\lambda+2\beta^2+2\gamma^2-\frac{q^2}{2}-q',
\end{eqnarray}
  where
  \begin{eqnarray}
  &&B_0=\Gamma_{456789}, B_1=\Gamma_{3456}, B_2=\Gamma_{3789},\\
&&  q=e^2\frac{d}{dy}\frac{1}{e^2}
\end{eqnarray}
and $'$ means $d/dy$. One can take $q,d\epsilon/dy$ and
  the parameters
  $\alpha, \beta, \gamma, s_i, t_j, u, v, w$ to be of first order, while $r, \tilde r$,
  the second derivatives of $e^2, \epsilon$, the first
  derivatives of the other parameters, and the quadratic expressions of the first order quantities
  to be of second
  order. Note that the first five equations come from the conditions
  that the first order variation of action under supersymmetry transformation
  vanish, while the vanishing of the second order variation leads to
  the last three equations. Introducing another parameter $\psi$, we
  set $\psi'=2\alpha$. We can solve the equation
  (\ref{epsiequation})
\begin{equation}
  \epsilon=(\cos\frac{\psi}{2}-\sin\frac{\psi}{2}B_0)\epsilon_0,\ \
  \ B_2\epsilon_0=\epsilon_0,\label{project-s}
\end{equation}
where $\epsilon_0$ is a constant spinor. The solution is
equivalent to impose the following projection condition
\begin{equation}
(\sin\psi B_1+\cos\psi B_2)\epsilon=\epsilon,\label{project}
\end{equation}
  which is the projection condition on $\epsilon$ representing half of
 supersymmetries. The $\epsilon$ also satisfy the requirement of closure
 of supersymmetry (\ref{susy}) if $\overline{\epsilon}_0\Gamma^3\tilde\epsilon_0=0$.
 The equations from (\ref{solution1}) to (\ref{solution2})
 are equivalent to the projection condition (\ref{project}). So it
 is easy to obtain the following results
\begin{eqnarray}
  &&\psi^{\prime}=2\alpha,\ \
  \beta=-\frac{\psi^{\prime}}{2\cos\psi},\ \
  \gamma=\frac{\psi^{\prime}}{2\sin\psi},\ \
  \frac{1}{e^2}=D\sin2\psi\\
  &&u=-4\alpha,\ \ v=-4\beta,\ \  w=-4\gamma,\ \ \theta=2\pi a+8\pi^2D\cos2\psi\\
  &&s_1=2\psi'\frac{\sin^2\psi}{\cos\psi},\ \
  s_2=2\psi'\sin\psi,\ \
  t_1=-2\psi'\cos\psi,\ \
  t_2=-2\psi'\frac{\cos^2\psi}{\sin\psi}.\label{st}
\end{eqnarray}
Since  $\theta$ and $1/e^2$ can be expressed in terms of the usual
complex coupling  parameter $\tau=\frac{\theta}{2\pi}+\frac{4\pi
i}{e^2}$ which
  takes values in the upper half plane, we have $\tau=a+4\pi
  De^{2i\psi}$.

  To obtain the vacuum structure of generalized Janus configurations, one may find
  the moduli space preserving full supersymmetries. However, the supersymmetry transformation parameter
   $\epsilon$  depends on $y$. It turns out to be more convenient to use the constant
   spinor $\epsilon_0$. All of the projection conditions on $\epsilon$
  can be reexpressed in terms of the projection conditions on the
  constant spinor
  $\epsilon_0$. We can use $\epsilon_0$ as the supersymmetry
  transformation parameter instead of $\epsilon$. The total supersymmetry transformation
  on gaugino is
\begin{equation}
   \delta\Psi=\frac{-1}{2}\left(\Gamma_{IJ}F^{IJ}
   +\Gamma^a X_a(s_1\Gamma_{456}+s_2\Gamma_{789})
   +\Gamma^p
   X_p(t_1\Gamma_{456}+t_2\Gamma_{789})\right)\epsilon.\label{Gagino-trans}
\end{equation}
Note that after using (\ref{st}) and
  the projection condition (\ref{project}), we have
\begin{eqnarray*}
&&\frac{1}{2}\left(\Gamma^a X_a\left(s_1\Gamma_{456}
  +s_2\Gamma_{789}\right)\right)\epsilon
  =\Gamma^{3a}X_a\frac{(\cos\psi)'}{\cos\psi}\epsilon\\
&&\frac{1}{2}\left(\Gamma^p
X_p(t_1\Gamma_{456}+t_2\Gamma_{789})\right)
\epsilon=\Gamma^{3p}X_p\frac{(\sin\psi)'}{\sin\psi}\epsilon,
\end{eqnarray*}
  where $a=4,5,6,\ \ p=7,8,9$. In order to
  make the expression simple, we denote
  \begin{equation}
  \overline{F}_{3a}=\frac{D_3(X_a\cos\psi)}{\cos\psi}
  ,\hspace{3ex}\overline{F}_{3p}=\frac{D_3(X_p\sin\psi)}{\sin\psi}.\end{equation}

Let us consider the moduli space of the theory. We can take
$A_{\mu}=0$ and six real scalars only depend on $x_3$. The vacuum
configurations preserve all supersymmetries. With the above ansatz,
the gaugino transformation (\ref{Gagino-trans}) becomes
\begin{eqnarray}
  \delta\Psi&=&-e^{-\psi B_0/2}\left(
  \left(-(X_a\cos\psi)'\Sigma^a+(X_p\sin\psi)'\Sigma^p
  -[X_a,X_p]\Sigma^a\Sigma^p\right)B_0\epsilon_0
  \right.\nonumber\\
  &&\left.+\left(\tan\psi(X_a\cos\psi)'\Sigma^a+\epsilon^{abc}[X_b,X_c]\Sigma^a/2+\cot\psi(X_p\sin\psi)'
  \Sigma^p+\epsilon^{pqr}[X_q,X_r]/2\right)\epsilon_0\right)\nonumber\\
  &=&0
\end{eqnarray}
where $\varepsilon^{456}=1,\ \varepsilon^{789}=1,\
2\Sigma^{a}=-\varepsilon^{abc}\Gamma_{bc},\
2\Sigma^{p}=-\varepsilon^{pqr}\Gamma_{qr}$.

  The Majorana-Weyl spinor in $16$ of $SO(1,9)$ can be decomposed as $V_8\otimes V_2$, where
$V_8$ transforms in the irreducible representation
$2\otimes2_X\otimes2_Y$ of $SO(1,2)\times SO(3)_X\times SO(3)_Y$ and
$V_2$ is the space acted by $SL(2,\mathbb{R})$ which is generated by
$B_0,\ B_1,\ B_2$. The $\Sigma^a$ are three generators of $SO(3)_X$
which acts on the space $2_X$ and $\Sigma^p$ are three generators of
$SO(3)_Y$ which acts on the space $2_Y$. The fifteen matrixes
$\Sigma^a,\ \Sigma^p,\ \Sigma^a\Sigma^p$ are anti-Hermitian and
traceless acting on $2_X\otimes2_Y$. As the trace of the product of
arbitrary two different matrixes vanishes, they are independent
matrixes acting on $2_X\otimes2_Y$. And since
$B_2\epsilon_0=\epsilon_0$ and $ B_2B_0\epsilon_0=-\epsilon_0 $, so
$\epsilon_0$ and $B_0\epsilon_0$ are two independent vectors in the
space $V_2$. Therefore the vanishing of the gaugino supersymmetry
transformation leads to
\begin{eqnarray}
  &&(X_a\cos\psi)'=0,\ \ \ \ (X_p\sin\psi)'=0\nonumber\\
  &&[X_a,X_b]=0,\ \ \ [X_p,X_q]=0,\ \ \ [X_a,X_p]=0,\label{moduli}
\end{eqnarray}
with the solution
\begin{equation}
  X_a=\frac{C_a}{\cos\psi},\ \ \ \ X_p=\frac{D_p}{\sin\psi},
\end{equation}
where $C_a,D_p$'s are constant matrices  commuting with each other.
This is the only vacuum we can have.

One can also study the vacuum directly from the energy functional
\begin{eqnarray}
  H&=&\int d^3x\frac{1}{e^2}\textrm{Tr}((\sin\psi \overline{F}_{34}-F_{56})^2
    +(\sin\psi \overline{F}_{35}-F_{64})^2+(\sin\psi
    \overline{F}_{36}-F_{45})^2\nonumber\\
    &&+(\cos\psi \overline{F}_{37}-F_{89})^2+(\cos\psi
    \overline{F}_{38}-F_{97})^2
    +(\cos\psi \overline{F}_{39}-F_{78})^2\nonumber\\
    &&+F_{ap}F^{ap})+\left(\frac{2}{e^2}\sin\psi\textrm{Tr}(X_4[X_5,X_6])
    +\frac{2}{e^2}\cos\psi\textrm{Tr}(X_7[X_8,X_9])\right)'\nonumber\\
    &&+\textrm{Tr}(\frac{1}{e^2}\psi'\tan\psi X_aX^a-\frac{1}{e^2}\psi'\cot\psi
    X_pX^p)'.\label{H}
\end{eqnarray}
The energy is bounded by the boundary term. If the boundary term
vanishes, the classical vacuum configurations satisfy the above
equations (\ref{moduli}) which we obtained from full supersymmetry
conditions of vacuum configurations. Thus we can conclude that there
is no nontrivial vacuum for Janus configuration with theta term.
This is quite different from the case without theta term, which has
a nontrivial vacuum characterized by a Nahm-like equation.

\section{ BPS solutions and Brane configurations}

   In this section, we try to obtain  the BPS
solutions which preserve part of the supersymmetries in generalized
Janus configuration. This require $\epsilon_0$ to satisfy extra
projection conditions which are also compatible with the original
condition $B_2\epsilon_0=\epsilon_0$. The possible projection
conditions for supersymmetric parameter $\epsilon_0$ are:
\begin{equation}
  \Gamma_{123p}\epsilon_0=\alpha\epsilon_0,\ \ \
  \Gamma^{0a}\epsilon_0=\beta\epsilon_0\label{Projection}
\end{equation}
where $\alpha=\pm 1,\beta=\pm 1$. Without losing generality, we just
set $p=7, a=4$ in the following discussion. In this case we have the
following identities:
\begin{eqnarray}
  \Gamma_{1289}\epsilon_0=-\alpha\epsilon_0,\ \ \
  \Gamma_{1256}\epsilon_0=\beta\epsilon_0,
  \ \ \ \Gamma_{3567}\epsilon_0=-\alpha\beta\epsilon_0,
  \ \ \ \Gamma_{5689}\epsilon_0=\alpha\beta\epsilon_0\\
  B_0\epsilon_0=-\Gamma_{0123}\epsilon_0=\beta\Gamma_{1234}\epsilon_0
  =-\Gamma_{3456}\epsilon_0=-\alpha\Gamma_{07}\epsilon_0=-\alpha\beta\Gamma_{47}\epsilon_0\\
  =\alpha\beta\Gamma_{3489}\epsilon_0=-\alpha\Gamma_{0389}\epsilon_0=\beta\Gamma_{0356}\epsilon_0
\end{eqnarray}
If we impose one projection condition in (\ref{Projection}) then we
get 1/2 BPS configurations. If imposing both conditions we obtain
1/4 BPS configurations. After multiplying a factor
$-(\cos\frac{\psi}{2}+\sin\frac{\psi}{2}B_0)$, the supersymmetry
transformation of the gaugino field  becomes
\begin{eqnarray*}
   &&\Gamma^{12}(F_{12}-F_{56}\Gamma_{1256}-F_{89}\Gamma_{1289}
    +\sin\psi\overline{F}_{34}\Gamma_{1234}-\overline{F}_{37}
    \cos\psi\Gamma_{1237})\epsilon_0\nonumber\\
   &&+\Gamma^{23}(F_{23}-\cos\psi F_{17}\Gamma_{1237}
      +\sin\psi F_{14}\Gamma_{1234}B_0)\epsilon_0\nonumber\\
   &&+\Gamma^{31}(\cos\psi F_{31}-\cos\psi F_{27}\Gamma_{1237}
      +\sin\psi F_{24}\Gamma_{1234}B_0)\epsilon_0\nonumber\\
   &&+(\cos\psi+\sin\psi
     B_0)\left[\Gamma^{15}(F_{15}+F_{26}\Gamma_{1256})\right.\nonumber\\
   &&\left.+\Gamma^{16}(F_{16}-F_{25}\Gamma_{1256})
     +\Gamma^{18}(F_{18}+F_{29}\Gamma_{1289})
     +\Gamma^{19}(F_{19}-F_{28}\Gamma_{1289})\right]\epsilon_0\nonumber\\
   &&+\Gamma^{01}(F_{01}+\cos\psi F_{14}\Gamma^{04}
     +\sin\psi F_{17}\Gamma_{07}B_0)\epsilon_0
     +\Gamma^{02}(F_{02}+\cos\psi F_{24}\Gamma^{04}\\
   &&+\sin\psi F_{27}\Gamma_{07}B_0)\epsilon_0
     +\Gamma^{03}(F_{03}+\cos\psi \overline{F}_{34}\Gamma^{04}
     +\sin\psi \overline{F}_{37}\Gamma_{07}B_0)\epsilon_0\\
   &&+\Gamma^{04}(\cos\psi F_{04}
     -\sin\psi F_{07}\Gamma_{47}B_0)\epsilon_0
     +\Gamma^{05}(\cos\psi F_{05}-F_{45}\Gamma^{04}\\
   &&-\sin\psi \overline{F}_{36}\Gamma_{0356}B_0)\epsilon_0
     +\Gamma^{06}(\cos\psi F_{06}-F_{46}\Gamma^{04}
     +\sin\psi \overline{F}_{35}\Gamma_{0356}B_0)\epsilon_0\\
   &&+\Gamma^{07}(\cos\psi F_{07}-F_{47}\Gamma^{04}
     +\sin\psi F_{04}\Gamma_{47}B_0)\epsilon_0
     +\Gamma^{08}(\cos\psi F_{08}-F_{48}\Gamma^{04}\\
   &&-\sin\psi \overline{F}_{39}\Gamma_{0389}B_0)\epsilon_0
     +\Gamma^{09}(\cos\psi F_{09}-F_{49}\Gamma^{04}
     +\sin\psi \overline{F}_{38}\Gamma_{0389}B_0)\epsilon_0\\
   &&+\Gamma^{58}(F_{58}+F_{69}\Gamma_{5689})\epsilon_0
     +\Gamma^{59}(F_{59}-F_{68}\Gamma_{5689})\epsilon_0
     +\Gamma^{35}(\cos\psi\overline{F}_{35}-F_{67}\Gamma_{3567}\\
   &&-\sin\psi F_{06}\Gamma_{0356}B_0)\epsilon_0
     +\Gamma^{36}(\cos\psi\overline{F}_{36}+F_{57}\Gamma_{3567}
     +\sin\psi F_{05}\Gamma_{0356}B_0)\epsilon_0\\
   &&+\Gamma^{38}(\cos\psi \overline{F}_{38}+F_{79}\Gamma^{3789}
     -\sin\psi F_{09}\Gamma_{0389}B_0)\epsilon_0\\
   &&+\Gamma^{39}(\cos\psi \overline{F}_{39}-F_{78}\Gamma^{3789}
     +\sin\psi F_{08}\Gamma_{0389}B_0)\epsilon_0
\end{eqnarray*}
The gaugino transformation should vanish for BPS configurations.
   Imposing projection condition (\ref{Projection}), $\delta\Psi=0$ if
   all terms vanish separately. This leads to the following nontrivial
   part of 1/4 BPS equations:
\begin{eqnarray}
&&F_{12}-\beta F_{56}+\alpha
   F_{89}+\beta\sin\psi\overline{F}_{34}-\alpha\cos\psi\overline{F}_{37}=0\nonumber\\
&&F_{23}+\beta\sin\psi F_{14}-\alpha\cos\psi F_{17}=0,\ \
  \ F_{31}+\beta\sin\psi F_{24}-\alpha\cos\psi F_{27}=0\nonumber\\
&&F_{18}-\alpha F_{29}=0,\ \ \ F_{19}+\alpha F_{28}=0\nonumber\\
&&F_{15}+\beta F_{26}=0,\ \ \ F_{16}-\beta F_{25}=0,\ \ \
   F_{59}-\alpha\beta F_{68}=0\nonumber\\
&&F_{58}+\alpha\beta F_{69}=0,\ \ \
 F_{01}+\alpha\sin\psi F_{17}+\beta\cos\psi F_{14}=0
   \nonumber\\
&&F_{02}+\alpha\sin\psi F_{27}+\beta\cos\psi F_{24}=0,\ \ \
  F_{03}+\alpha\sin\psi
  \overline{F}_{37}+\beta\cos\psi\overline{F}_{34}=0\nonumber\\
&&\cos\psi F_{04}+\alpha\beta\sin\psi F_{07}=0,\ \ \
  \cos\psi F_{05}-\beta F_{45}+\beta\sin\psi \overline{F}_{36}=0\nonumber\\
&&\cos\psi F_{06}-\beta F_{46}-\beta\sin\psi \overline{F}_{35}=0,\ \
   \ \cos\psi F_{07}-\beta F_{47}-\alpha\beta\sin\psi F_{04}=0\nonumber\\
&&\cos\psi F_{08}-\beta F_{48}-\alpha\sin\psi \overline{F}_{39}=0,\
  \ \  \cos\psi F_{09}-\beta F_{49}+\alpha\sin\psi
  \overline{F}_{38}=0\nonumber\\
&&\cos\psi \overline{F}_{35}+\beta\sin\psi F_{06}+\alpha\beta
  F_{67}=0,\ \ \ \cos\psi \overline{F}_{36}-\beta\sin\psi F_{05}
  -\alpha\beta F_{57}=0\nonumber\\
&&\cos\psi \overline{F}_{38}-\alpha\sin\psi F_{09}+F_{79}=0,\ \ \
  \cos\psi \overline{F}_{39}+\alpha\sin\psi F_{08}-F_{78}=0\label{bps}
\end{eqnarray}
  These equations contain unknown parameter $\psi$ which
  depends on $y$, and seem to be impossible to solve.

 It is easier to deal with half BPS configurations. The
nontrivial part of half BPS equation with one of the projection
conditions
  $\Gamma_{1237}\epsilon_0=\alpha\epsilon_0$ and $\beta=0$ is made of
\begin{eqnarray}
&&F_{12}+\alpha
   F_{89}-\alpha\cos\psi\overline{F}_{37}=0,\ \ \
  F_{23}-\alpha\cos\psi F_{17}=0\nonumber\\
&& F_{31}-\alpha\cos\psi F_{27}=0,\ \ \
   F_{01}+\alpha\sin\psi F_{17}=0\nonumber\\
&&F_{02}+\alpha\sin\psi F_{27}=0,\ \ \
  F_{03}+\alpha\sin\psi\overline{F}_{37}=0\nonumber\\
&&F_{18}-\alpha F_{29}=0,\ \ \ F_{19}+\alpha F_{28}=0\nonumber\\
&&\cos\psi F_{08}-\alpha\sin\psi \overline{F}_{39}=0,\
  \ \ \cos\psi \overline{F}_{38}-\alpha\sin\psi F_{09}+F_{79}=0\nonumber\\
&&\cos\psi F_{09}+\alpha\sin\psi
  \overline{F}_{38}=0,\ \ \
  \cos\psi \overline{F}_{39}+\alpha\sin\psi F_{08}-F_{78}=0.
\end{eqnarray}
The last two lines of the above equations could be reduced to
\begin{eqnarray}
\overline{F}_{39}=\cos\psi F_{78}& &F_{08}=\alpha\sin\psi F_{78}
\\
\overline{F}_{38}=-\cos\psi F_{79}& &F_{09}=\alpha\sin\psi F_{79}.
\end{eqnarray}
 One can simplify the equations further by let $X_8=X_9=0$, then the first six equations are the
dyon equation when $\psi$ is a constant. Although generically
$\psi$ is not a constant in the Janus field theory, the above
equations could be organized as
 \begin{eqnarray}
 B_i=\alpha\cos\psi D_i  X_7,& i=1,2,\hspace{3ex}&B_3=\alpha\cos\psi\frac{D_3\overline{X}_7}{\sin\psi}\nonumber\\
  E_i=-\alpha \sin\psi D_i X_7,& i=1,2,\hspace{3ex}&E_3=-\alpha \sin\psi\frac{D_3\overline{X}_7}{\sin\psi}
 \label{bpsX7}
 \end{eqnarray}
where $\overline{X}_7=\sin\psi X_7,\ E_i=F_{0i},\
B_i=\frac{1}{2}\epsilon^{ijk}F_{jk}$\footnote{This is notation for
the magnetic field strength without confusing with the previous
notation for gamma matrix products.}. One may take them as modified
equations for dyons, as we will show soon. They are quiet different
from the monopole equations in \cite{KKL}. It would be
interesting to solve these equations. We will discuss the solution in the sharp interface case.%It can be understood from
%brane configuration.

The trivial part of the BPS equation involves the equations on
$X_{4,5,6}$, which requires them to be constant. For simplicity, we
set $X_{4,5,6}=0$.

   Let us consider the energy functional in the case that the six
adjoint scalars  except $X_7$ vanish. The energy functional takes
the following form
\begin{eqnarray}
   H&=&-\int d^3x\frac{1}{e^2}\textrm{Tr}((B_i-\alpha\cot\psi
   D_i\overline{X}_{7})^2
     +(E_i+\alpha D_i\overline{X}_{7})^2)\nonumber\\
     &&+\alpha\int d^3x
     \partial_i\textrm{Tr}(\frac{2}{e^2}(\cot\psi B_i\overline{X}_7-
     E_i\overline{X}_7))
\end{eqnarray}
 To get
it, we have used the Gauss law
\begin{equation}\label{gauss}
   D_i(\frac{1}{e^2}E_i)+2\frac{\psi'}{e^2}B_3=0.
\end{equation}
It is obvious that the energy functional is in consistent with the
BPS equation we obtained from supersymmetry analysis, if  the
boundary term is  ignored.

%   However, we can obtain the ratio between the electric charge and the magnetic charge
% from the BPS equations even if we don't know the
%specific solutions.
Since it is hard to solve the half-BPS equations, it is useful to
recall the BPS soliton in usual field theory.  For simplicity, let
us assume the gauge group to be $SU(2)$. The adjoint scalar is
written as $\phi$ taking vacuum expectation value
$\phi^a\phi^a=e^2v^2$ in the spatial infinity where $SU(2)$ is
broken to $U(1)$. In the traditional field theory, the coupling is a
constant and the magnetic charge $Q_m$ is related to the winding
number
\begin{eqnarray}
  Q_m&=&2\int_{s^2_{\inf}} d S^i \frac{1}{e^2 v}\textrm{Tr}(\phi B_i)=\frac{4\pi
  n_m}{e}\label{qm}
\end{eqnarray}
 where $n_m$ is the winding number of scalar field configuration. However when the coupling
 is spatially varying, the relation between
magnetic charge and the winding number is not clear.  For the
electric charge,
\begin{eqnarray}
   Q_e=-2\int_{s^2_{\inf}} d S^i
   \textrm{Tr}\frac{1}{e^2 v}E_i\phi,
\end{eqnarray}
we have to take into account of the Witten effect \cite{Wi} in the
presence of the theta term. The generator that generates the gauge
transformations around the direction $\phi^a$ is $\delta
A_{\mu}^a=(1/ev)D_{\mu}\phi^a$, whose corresponding Noether charge
is
\begin{eqnarray}
   n_e&=&\int d^3x \frac{\partial \pounds}{\partial(\partial_0A^a_{\mu})}\delta
   A^a_{\mu}\nonumber\\
   &=&\int_{s^2_{\inf}} d S^i
   \textrm{Tr}(-\frac{2}{e^3v}E_i\phi+\frac{\theta}{4\pi^2ev}B_i\phi)\nonumber\\
   &=&\frac{Q_e}{e}+\frac{\theta e}{8\pi^2 }Q_m \label{qe}
\end{eqnarray}
 by using the Gauss Law.  $n_e$ must be an integer, since $2\pi$
 rotation is not a transformation and leaves the state invariant,
 giving $e^{i2\pi n_e}=1$.

 In our case, we can identify $X_7$ as the scalar field $\phi$
 above. Then using the corresponding BPS equations (\ref{bpsX7}) and
$1/e^2=D\sin2\psi,\theta=2\pi(a+4\pi D\cos2\psi)$,
 one can find
\begin{equation}\label{nem}
   n_e/n_m=a+4\pi D.
\end{equation}
We will show that the similar relation also appears in the
discussion of brane configuration of BPS solition as
$(p,q)$-string. This  suggest $n_e$ and $n_m$ should be identified
as the charges  carried by $(p,q)$ string on D3-brane worldvolume.
%So it is consistent for the Janus field and the correspondent
%brane construction that there is actually a constraint on the
%ratio of the charges for BPS configuration.

For Janus configuration with spatially varying coupling and theta
term, one has to carefully define the electric and the magnetic
charge of the BPS soliton. Let us take the following definition,
which could be reduced to the usual ones consistently:
\begin{eqnarray}
Q_m=2\int d^3x\partial_i \textrm{Tr}(\frac{1}{e^2v}B_iX_7),
\hspace{3ex} Q_e=-2\int d^3x\partial_i
\textrm{Tr}(\frac{1}{e^2v}E_iX_7).
\end{eqnarray}
%With the given the topological charge and the Noether charge,

The energy of half BPS soliton solution is bounded by
\begin{equation}
   H\geqslant |\int d^3 x
   \partial_i\textrm{Tr}(\frac{2}{e^2}(\cos\psi B_iX_7-
     \sin\psi E_iX_7))|
\end{equation}
  with equality if and only if  the
BPS equations are obeyed. This inequality looks different from the
usual BPS relation between the mass and the charges. However, if
we consider a simple case with $\psi$ being a constant, then we
have
\begin{eqnarray}
M&\geq &v|\cos\psi Q_m +\sin\psi Q_e| \nonumber \\
 &\geq &v \sqrt{Q_m^2+Q_e^2}.
 \end{eqnarray}
In the above relation the equality is saturated if the BPS equation
is satisfied and $\tan\psi=Q_e/Q_m$ which does hold taking into
account of the relations (\ref{qe},\ref{qm},\ref{nem}) and the
expression of $e^2$ and $\theta$ in terms of $\psi$. This suggests
that the half BPS solution is actually a dyon. For a dyon with
charges $(p,q)$, $M$ is proportional to $|p+q\tau|$ with $\tau$
being a complex coupling constant, and
\begin{equation}\label{psi}
\psi+\frac{1}{2}\pi=\arg(p+q\tau).
\end{equation}
 This sounds somehow strange
since $\psi$ is an arbitrary function in Janus configuration.
However, as we will show in the discussion on brane realization of
BPS soliton, in order to have half BPS configuration, the possible
$(p,q)$ string is very restricted and indeed (\ref{psi}) must be
respected.

Here seems a puzzle. When $\psi$ being a constant, the Lagrangian
of Janus configuration reduces simply to the one of ${\cal N}=4$
super-Yang-Mills theory. And there should be various possible
dyonic solutions in the theory. At first sight this is in
contradiction with the result we just obtained. However, note that
even when $\psi$ being a constant, we have extra projection
condition on supersymmetry parameter and the field theory has
actually half of the original supersymmetries. The extra
projection condition leads to stringent constraints on the
possible BPS solitonic solutions. In the brane picture, the extra
projection condition comes from the presence of 5-brane system.
The constancy of $\psi$ along the whole $y$ corresponds to the
case the $\psi$ being the same on two sides of 5-brane system. In
this case, the explicit solution of BPS soliton solution is the
same as the one in ${\cal N}=4$ SYM with the dyon charges being
relatively fixed.

Next let us consider a little more complicated case. We assume
that $\psi$ take different values on two sides of 5-brane locating
at $y=0$. In this so-called sharp interface case, $\psi$ changes
from one constant value to another one at the interface, namely
\begin{equation}
 \psi(y)=\left\{\begin{array}{ll}
  \psi_1, \ y>0\\
  \psi_2, \ y<0.
  \end{array}\right.
\end{equation}
   Similar to \cite{KKL} one can still solve the BPS equations in the abelian limit that
the nonabelian core size vanishes. In the abelian limit, the
question is simplified to the one in traditional electrodynamics.
However, one has to be careful since we are discussing the dyon
which induce both point-like electric and magnetic sources. In fact,
from BPS equation, the fact $B_i/E_i=-\cot\psi$ suggests that $B_i$
and $E_i$ cannot be both continuous across the interface since
$\psi$ is different on two sides.  In the following discussion, we
just focus on the magnetic field and the electric field is given by
$E_i=-\tan\psi B_i$. For a single dyon with one unit of magnetic
charge at $y=y_0>0$, we have\footnote{For a realistic dyon with
charge $(p,q)$, the electric and magnetic field strength is simply
the multiple of the ones for one unit charge in the Abelian limit.
Therefore, we just focus on the case with one unit charge. }
\begin{equation}
 B_i=\cot\psi D_i\overline{X}_7=i\sigma_3/2\left\{\begin{array}{ll}
  \frac{(x_1,x_2,y-y_0)}{r_+^3}+\frac{\cot^2\psi_1-\cot^2\psi_2}
  {\cot^2\psi_1+\cot^2\psi_2}\frac{(x_1,x_2,y+y_0)}{r_-^3}, \ y>0\\
  \frac{2\cot^2\psi_2}
  {\cot^2\psi_1+\cot^2\psi_2}\frac{(x_1,x_2,y-y_0)}{r_+^3}, \  y<0
  \end{array}\right.
\end{equation}
where $r_{\pm}=x_1^2+x_2^2+(y\mp y_0)^2$.  The scalar field is
\begin{equation}
 {X}_7=i\sigma_3/2\left\{\begin{array}{ll}
  \frac{1}{\sin\psi_1}(\tilde v-\frac{1}{\cot^2\psi_1r_+}-\frac{\cot^2\psi_1-\cot^2\psi_2}
  {\cot^2\psi_1(\cot^2\psi_1+\cot^2\psi_2)}\frac{1}{r_-}), \ y>0\\
   \frac{1}{\sin\psi_2}(\tilde v-\frac{2}
  {\cot^2\psi_1+\cot^2\psi_2}\frac{1}{r_+}), \  y<0
  \end{array}\right.
\end{equation}
where $\tilde v=\sqrt{\frac{\sin\psi_1}{2D\cos\psi_1}} v$ and $v$ is
a constant being related to $X_7/e$ at positive infinity. It is easy
to see that the total magnetic flux is $4\pi$ at the spacial
infinity.  And the charges of the dyon are
\begin{eqnarray}
  Q_m=\frac{4\pi
  (\cos\psi_1\cot^2\psi_1+\cos\psi_2\cot^2\psi_2)}{\cot\psi_1^2+\cot^2\psi_2}
  \sqrt{\frac{2D\sin\psi_1}{\cos\psi_1}}\\
   Q_e=\frac{4\pi
  (\sin\psi_1\cot^2\psi_1+\sin\psi_2\cot^2\psi_2)}{\cot\psi_1^2+\cot^2\psi_2}
  \sqrt{\frac{2D\sin\psi_1}{\cos\psi_1}}
\end{eqnarray}
when $\psi_1=\psi_2$ we have $Q_m=4\pi/e$ which is the charge of a
single monopole. To obtain the above solution, one needs to take
into continuity condition on various fields. Here we take $F_{01},
F_{02}, F_{12}$ and $\overline{X}_7$ to be continuous at the
interface. Obviously the scalar field $X_7$ is not continuous.

 Another simplification is to let $A_i=0,\ A_0=\overline{X}_7$ and $X_{7,8,9}$
depend only on $y$. This leads to the following equations
 \begin{eqnarray}
 F_{89}&=&\cos\psi\overline{F}_{37}\\
 F_{97}&=&\frac{1}{\cos\psi}\overline{F}_{38}\\
 F_{78}&=&\frac{1}{\cos\psi}\overline{F}_{39}. \label{Nahmlike}
 \end{eqnarray}
 One can also obtain the energy functional in this case
\begin{eqnarray}
  H&=&-\int
  d^3x\frac{1}{e^2}\textrm{Tr}(F_{89}-\cos\psi\overline{F}_{37})^2
  +(F_{03}+\sin\psi\overline{F}_{37})^2\nonumber\\
  &&+(\cos\psi F_{08}-\sin\psi
  \overline{F}_{39})^2
  +(\cos\psi \overline{F}_{38}
  -\sin\psi F_{09}+F_{79})^2\nonumber\\
  &&+(\cos\psi F_{09}+\sin\psi
  \overline{F}_{38})^2+(\cos\psi \overline{F}_{39}+\sin\psi
  F_{08}-F_{78})^2\nonumber\\
  &&+(\frac{1}{e^2}(2\cos\psi\textrm{Tr}(X_7[X_8,X_9]-\psi'\cot\psi
    \textrm{Tr}X_pX^p-2F_{03}\overline{X}_7))'
\end{eqnarray}
by using the following equation of motion
\begin{equation}
  D_3(\frac{1}{e^2}F_{03})+[X_8,\frac{1}{e^2}F_{08}]+[X_9,\frac{1}{e^2}F_{09}]=0.
\end{equation}
Similarly it is in consistent with the BPS equations. One can also
obtain the $1/4$ BPS equation from the energy functional analysis
which is similar to the above case. It contains some boundary terms
and square terms which are left hand of the equations in
(\ref{bps}).

To solve the BPS equations, without losing generality, we can assume
the gauge group to be $SU(2)$ and make the following ansatz,
 \begin{equation}
 X_7=-if_1(y)\sigma_1/2, \hspace{3ex}X_8=-if_2(y)\sigma_2/2,
 \hspace{3ex}X_9=-if_3(y)\sigma_3/2,
 \end{equation}
 with $\sigma_i$'s being Pauli matrices.   The above set of equations
could be reduced to
\begin{equation}
 f_2f_3=\frac{\cos\psi\partial_y(f_1\sin\psi)}{\sin\psi},
 \hspace{3ex}f_1f_3=\frac{\cos\psi\partial_y(f_2\sin\psi)}{\sin\psi},
 \hspace{3ex}f_1f_2=\frac{\partial_y(f_3\sin\psi)}{\cos\psi\sin\psi}.
 \end{equation}
 If $\psi$ is a constant, the above equation could be solved by a
 proper rescaling of $f_3(y)$. The solutions are
  \begin{eqnarray}
  f_1(y;k,F,y_0)&=&-\frac{Fcn_k(F(y-y_0))}{sn_k(F(y-y_0))}\nonumber
  \\
 f_2(y;k,F,y_0)&=&-\frac{Fdn_k(F(y-y_0))}{sn_k(F(y-y_0))}\nonumber
  \\
 f_3(y;k,F,y_0)&=&-\frac{F\cos\psi}{sn_k(F(y-y_0))}\nonumber
 \end{eqnarray}
where $sn_k,cn_k,dn_k$ are Jacobi elliptic functions with $k$ being
elliptic modulus and $F\geq 0, y_0$ are arbitrary constants. However
if $\psi$ is an arbitrary function of $y$, the equations
(\ref{Nahmlike}) become very difficult to solve.

  Another projection condition
  $\Gamma^{04}\epsilon_0=\beta\epsilon_0$ and $\alpha=0$ leads to another
  half BPS configurations
\begin{eqnarray}
&&F_{12}-\beta F_{56}+\beta\sin\psi\overline{F}_{34}=0,\ \ \
  F_{23}+\beta\sin\psi F_{14}=0\nonumber\\
&&F_{31}+\beta\sin\psi F_{24}=0,\ \ \
  F_{01}+\beta\cos\psi F_{14}=0\nonumber\\
&&F_{02}+\beta\cos\psi F_{24}=0,\ \ \
  F_{03}+\beta\cos\psi\overline{F}_{34}=0\nonumber\\
&&F_{15}+\beta F_{26}=0,\ \ \ F_{16}-\beta F_{25}=0\nonumber\\
 &&\cos\psi
F_{05}-\beta F_{45} +\beta \sin\psi \overline{F}_{36}=0, \ \ \
\cos\psi \overline{F}_{35}+\beta\sin\psi F_{06}=0
\nonumber\\
&&\cos\psi F_{06}-\beta F_{46}-\beta\sin\psi \overline{F}_{35}=0,\
\ \ \cos\psi \overline{F}_{36}-\beta\sin\psi F_{05}=0.\label{04}
\end{eqnarray}
 The discussion is very similar to the former case.

 Let us return to the 1/4-BPS equations. Since the projection
 conditions only involve two directions $4$ and $7$, one may
 simplify the discussion by set $X_5=X_6=X_8=X_9=0$ such that the
 equations could be rewritten as
 \begin{eqnarray}
 D_iX_4&=&-\beta(\cos\psi E_i+\sin\psi B_i),\hspace{3ex} i=1,2 \nonumber \\
 D_iX_7&=&\alpha(-\sin\psi E_i+\cos\psi B_i), \hspace{3ex} i=1,2 \nonumber \\
 \frac{D_3(X_4\sin\psi)}{\sin\psi}&=&-\beta(\cos\psi E_3+\sin\psi
 B_3),\nonumber \\
 \frac{D_3(X_7\sin\psi)}{\sin\psi}&=&\alpha(-\sin\psi E_3+\cos\psi
 B_3),\nonumber \\
 D_0X_4&=&D_0X_7=0, \hspace{3ex}[X_4,X_7]=0.
 \end{eqnarray}
When $\psi$ is simply a constant, the above equations are the ones
for BPS string junctions in SYM\cite{BEG,HHS}. Therefore the above
equations could be taken as the ones for BPS string junction in
Janus configuration. It would be interesting to solve these
equations.

   Next, we will analyze the corresponding brane
configurations of the BPS solutions. The brane construction of the
theory is D3-brane ending on five-brane. D3-brane extends along the
0123 directions. There may be two groups of five-branes. One of the
($p,q$) five-brane extends along 012456 directions and the other one
($p',q'$) extends along 012789. If one takes $\epsilon_1$ and
$\epsilon_2$ as the supersymmetry parameters of the left and right
move modes in type IIB theory, then the unbroken supersymmetry for
D3-brane is just
\begin{eqnarray}
  \epsilon_2=\Gamma_{0123}\epsilon_1 \label{d3-pro}
\end{eqnarray}

  The $\epsilon_1$ is identical to $\epsilon$ in the above Janus field
theory. Using (\ref{project-s},\ref{project}), we also have
\begin{equation}
  \epsilon_1=(\cos\frac{\psi}{2}-\sin\frac{\psi}{2}B_0)\epsilon_0,\
  \ \ (\sin\psi B_1+\cos\psi B_2)\epsilon_1=\epsilon_1 \label{1-pro}
\end{equation}
As we have argued before, the projection condition on $\epsilon_1$
can be transformed to the projection condition on $\epsilon_0$. We
can write it as $\Gamma'\epsilon_0=\epsilon_0$. To be compatible
with the original projection condition $B_2\epsilon_0=\epsilon_0$
asks that $\Gamma'$ and $B_2$ must commute.
%We can also impose more projection conditions on $\epsilon_0$ only
%if each of the projection matrixes commute. It is enough for
%obtaining BPS equations.
But if we want to know the brane configurations of the corresponding
BPS configurations, we should know the projection condition on
$\epsilon_1$. Let us consider a ($p,q$)-string extended along 0m
directions. The supersymmetry condition for $(p,q)$-string is
\begin{equation}
   \epsilon_1=-\Gamma(\cos t\epsilon_1-\sin t\epsilon_2)
\end{equation}
 where $\Gamma=\Gamma_{0m}$ and $ t=\textrm{arg}(q\tau+p)$.
  According to (\ref{d3-pro}), it is equivalent to
$\epsilon_1=-\Gamma(\cos t\epsilon_1+\sin tB_0\epsilon_1)= -\Gamma
e^{tB_0}\epsilon_1$. Using (\ref{1-pro}), we express it in terms
of $\epsilon_0$ as follows
\begin{equation}
   \epsilon_0=-e^{\psi B_0/2}\Gamma e^{tB_0}e^{-\psi
   B_0/2}\epsilon_0.
\end{equation}

If the ($p,q$)-string  extends along 1, 2 or 3 directions, it will
be dissolved in D3 branes and form a bound state\cite{KNP} whose
supersymmetry conditions are different from (\ref{d3-pro}). In this
case, one has to find a noncommutative field theory for Janus
configuration\cite{HHS}. For the ($p,q$)-string extending along
other directions, we have $\Gamma B_0=-B_0\Gamma$, and then
 \begin{equation}
\epsilon_0=-\Gamma
e^{(t-\psi)B_0}\epsilon_0
  =-(\cos (t-\psi)\Gamma+\sin (t-\psi)\Gamma B_0)\epsilon_0.
  \end{equation}
 The   compatible condition leads to
\begin{equation}
  \cos(t-\psi)=0,\hspace{5ex} \mbox{if $\Gamma
  B_2=-B_2\Gamma$}\label{ga07}
 \end{equation}
 with the solutions
 \begin{equation}
 \left\{\begin{array}{ll}
   t=\psi+\pi/2,& -\Gamma B_0\epsilon_0=\epsilon_0\\
   t=\psi-\pi/2,& \Gamma B_0\epsilon_0=\epsilon_0
   \end{array}\right.\label{sol07}
 \end{equation}
    or
\begin{equation}
  \sin(t-\psi)=0,\hspace{5ex} \mbox{if $\Gamma
  B_2=B_2\Gamma$,}\label{ga04}
 \end{equation}
 with the solutions
\begin{equation}
 \left\{\begin{array}{ll}
  t=\psi,& \Gamma\epsilon_0=-\epsilon_0\\
  t=\psi+\pi, & \Gamma\epsilon_0=\epsilon_0
  \end{array}\right.\label{sol04}
\end{equation}

   The relation of $t$ and $\psi$ can determine $p/q$ in terms of $a, D$ or
   vice versa. For example $t=\psi$ is equivalent to
\begin{equation}
   \frac{\sin\psi}{\cos\psi}=\frac{\textrm{Im}(p+q\tau)}{\textrm{Re}(p+q\tau)},
\end{equation}
which leads to $p/q=4\pi D-a$ by
   using $\tau=a+4\pi De^{2i\psi}$. Similarly $t=\psi+\pi/2$ leads to $p/q=-a-4\pi
   D$. %Since existing  constraints on charge it look like the Janus
   %field theory more generalized than the correspondent brane
   %configuration.

    The above analysis is for the general case.
   Now let us be more specific.  For
($p,q$)-string extending along 04  we have $\Gamma=\Gamma_{04}$.
Since $\Gamma_{04}B_2=B_2\Gamma_{04}$ satisfying (\ref{ga04}),  we
have the corresponding solution (\ref{sol04}) that
$\Gamma^{04}\epsilon_0=\epsilon_0$ for $t=\psi$ or
$\Gamma^{04}\epsilon_0=-\epsilon_0$ for $t=\psi+\pi$. The constraint
for the charge is
\begin{equation}\label{04Da}
p/q=4\pi D-a. \end{equation} Here the projection condition is
exactly the same one in (\ref{Projection}) for $a=4$ in
generalized Janus configuration. Thus the half BPS solution
(\ref{04}) we obtained above could be realized by the brane
configuration with ($p,q$)-string extending along 04. The
discussion  for $(p,q)$-strings along 05 and 06 cases is similar.

The thing is actually a little subtler here. In the brane picture,
the integration constants $a$ and $D$ are determined by the
background branes. For the generalized Janus configuration, one can
consistently add 5-branes along 012456 or along 012789 or
both\cite{GW1}. To be consistent with (\ref{04Da}), generically only
5-branes extending along 012456 are allowable. This brane picture
may help us to understand the equations in (\ref{04}). For example,
 the
fact that $(p,q)$-string along 04 looks like dyon from the point of
view of D3-brane extending along 0123 explains the modified dyon
equations in (\ref{04}). And the fact that $(p,q)$-string realize
the instantons in transverse directions 1256 of 5-branes is encoded
in the fourth line of (\ref{04}).

 On the other hand, for the strings along 07 case the
projection condition is $\Gamma_{1237}\epsilon_0=\pm \epsilon_0$ as
$\Gamma_{07}B_2=-B_2\Gamma_{07}$ and the charges has to satisfy
$p/q=-a-4\pi D$, which asks the 5-branes to lie along 012789
consistently. In this case, the parameter $t=\psi+\pi/2$ matches
exactly with the relation (49) and the charge ratio is reminiscent
of (45). This is exactly in match with the half-BPS solutions in
generalized Janus configuration obtained by imposing one of the
projection condition in (\ref{Projection}) with $p=7$. The string
orthogonal to D3 branes worldvolume realize a dyon in D3 branes. For
the strings lie along 08 or 09, we have the same picture.

%  However, it seems like to be some insistent for the brane analysis
%and Janus field theory. From BPS equations (\ref{bps}), taking the
%scalars vanishes except for $x_4,X_7$ and $\psi$ varying very slow,
%one can think that there are dyons \cite{HHS} correspondent to
%string network \cite{BEG}. But we can not find string network from
%the brane configurations.

\section{Less supersymmetry Janus Configurations With Theta Angle}

  Less supersymmetry Janus Yang-Mills theory without theta angle have been considered
  in \cite{HEG0613}\cite{KKL}. We can use the similar technique
  to investigate the case with theta angle. We can impose additional
  project condition to the susy parameter $\epsilon_0$ as long as the conditions are all consistent.
   In order to make the
  expression simple, we use the following notation:
\begin{eqnarray}
   &&B_{01}=\Gamma_{3456},\ \ B_{11}=\Gamma_{3489},\ \
   B_{21}=\Gamma_{3597},\ \ B_{31}=\Gamma_{3678},\\
   &&B_{i2}=B_0 B_{i1}, \ \ \ i=0,1,2,3
\end{eqnarray}
They satisfy the relations
$B_0^2=-1,B_{i1}^2=B_{i2}^2=1,B_0B_{i1}=-B_{i1}B_0$.  The projection
conditions for 1/4 supersymmetry configuration which have only two
supercharges are
\begin{equation}
  B_{i2}\epsilon_0=h_i\epsilon_0,
\end{equation}
where $h_i=\pm 1,i=0,1,2,3$. The compatible conditions require
$h_0h_1h_2h_3=-1$. In fact there are only three independent
constraints. They are equivalent to
\begin{equation}
 (\sin\psi B_{i1}+\cos\psi
B_{i2})\epsilon=h_i\epsilon.\label{identity1}
\end{equation}
   Similar to \cite{GW1} we can also analyze the brane configuration of the
fivebranes  that lead to the above  supersymmetries. For
($p,q$)-fivebrane extending along the 012456 directions imposes a
constraint
\begin{equation}
   \epsilon_1=-\Gamma_{012456}(\sin t\epsilon_1+\cos t\epsilon_2).
\end{equation}
  If existing both types of branes, then we have
\begin{equation}
   \epsilon_1=-(-\cos t B_{11}+\sin tB_{12})\epsilon_1,
\end{equation}
where $t=\textrm{arg}(q+p\tau)$. Compared with (\ref{identity1}), we
find $t=\psi\pm\pi/2$. It leads to a constraint on the charges that
is $q/p=-a-4\pi D$. Other projection conditions correspond to
($p,q$)-fivebranes extended along 012489, 012597 or 012678. Their
constraint on charges are the same so that they have the same charge
for fivebrane extended along these directions. However the fivebrane
also can extend along 012789, 012567, 012648 or 012459 and they have
the requirement $t=\psi$ or $t=\psi+\pi$ which lead to the same
condition of the charge $q/p=-a+4\pi D$.

The perturbed supersymmetry transformation could be
\begin{eqnarray}
     \delta_1\Psi&=&\frac{-1}{2}C_i(\Gamma^{a_i}X_{a_i}(s_{i1}\Gamma_{3}B_{i1}+s_{i2}\Gamma_{3}B_{i2})
     +\Gamma^{p_i}X_{p_i}(t_{i1}\Gamma_{3}B_{i1}+t_{i2}\Gamma_{3}B_{i2}))\epsilon,
\end{eqnarray}
where
$a_0=4,5,6,a_1=4,8,9,a_2=5,9,7,a_3=6,7,8,p_0=7,8,9,p_1=5,6,7,p_2=6,4,8,p_3=4,5,9$
and $C_i$'s are constants. The ansatz for perturbed action is
\begin{eqnarray}
   I^{\prime}&=& \int d^4x\frac{i}{e^2}\textrm{Tr}\overline{\Psi}\left(\alpha\Gamma_{012}
   +C_i(\beta_i\Gamma_{3}B_{i1}+\gamma_i\Gamma_{3}B_{i2})\right)\Psi,\\
   I^{\prime\prime}&=&\int d^4x\frac{1}{e^2}(u\varepsilon^{\mu\nu\lambda}
   \textrm{Tr}(A_{\mu}\partial_{\nu}A_{\lambda}+\frac{2}{3}A_{\mu}A_{\nu}A_{\lambda})\nonumber\\
   &&+\sum_{i=0}^{3}(C_i\frac{v_i}{3}\varepsilon^{a_i b_i
   c_i}\textrm{Tr}X_{a_i}[X_{b_i},X_{c_i}]
   +C_i\frac{w_i}{3}\varepsilon^{p_i q_i
   r_i}\textrm{Tr}X_{p_i}[X_{q_i},X_{r_i}])),\\
   I^{\prime\prime\prime}&=&\int
   d^4x\textrm{Tr}(\frac{1}{e^2}r_{mn}X_mX_n),
\end{eqnarray}
where $\varepsilon$ are antisymmetric tensors  normalized as
$\varepsilon^{456}=\varepsilon^{489}=\varepsilon^{597}=\varepsilon^{678}=
\varepsilon^{789}=\varepsilon^{567}=\varepsilon^{648}=\varepsilon^{459}=1$.
There are undetermined parameters $\alpha, \beta_i,\gamma_i,
u,v_i,w_i, r_{mn}$ in the perturbed action, which should be
determined by supersymmetry. Since the zeroth order and first order
terms under supersymmetry variation are linear in $C_0,C_1,C_2,C_3$,
the terms containing $C_i$ vanish separately. Then the equations for
the parameters are similar to
(\ref{epsiequation})--(\ref{solution3}), where
$B_1,B_2,\beta,\gamma,v,w$ become
$B_{i1},B_{i2},\beta_i,\gamma_i,v_i,w_i$. So we have similar
solutions for the undetermined parameters
\begin{eqnarray}
  &&\psi^{\prime}=2\alpha,\ \
  \beta_i=-h_i\frac{\psi^{\prime}}{2\cos\psi},\ \
  \gamma_i=h_i\frac{\psi^{\prime}}{2\sin\psi},\\
  &&u=-4\alpha,\ \ \ v_i=-4\beta_i,\ \ \ w_i=-4\gamma_i,\ \
  \frac{1}{e^2}=D\sin2\psi,\\
  &&s_{i1}=2h_i\psi'\frac{\sin^2\psi}{\cos\psi},\ \
  s_{i2}=2h_i\psi'\sin\psi,\label{si}\\
  &&t_{i1}=-2h_i\psi'\cos\psi,\ \
  t_{i2}=-2h_i\psi'\frac{\cos^2\psi}{\sin\psi}.\label{ti}
\end{eqnarray}
However there is an additional constraint for $C_i$ that is $\sum_i
C_i=1$.

The second order variations of the action come from three parts. The
first part is the perturbed modified supersymmetry transformation of
$I'$, which will be denoted as $\delta_1I'$. The second one is from
$\delta_1I$ which contains both the first order and the second order
variation parts. We denote the second order variation part as
$\delta_1I|_2$. The last one is from the unperturbed supersymmetry
variation of $I^{\prime\prime\prime}$. Their expressions are the
following
\begin{eqnarray}
  \delta_1I^{\prime}&=&-i\int d^4x\frac{1}{e^2}\textrm{Tr}\overline{\epsilon}C_i\left(
  (s_{i1}B_{i1}+s_{i2}B_{i2})\Gamma^{a_i}X_{a_i}
   +(t_{i1}B_{i1}+t_{i2}B_{i2})\Gamma^{p_i}X_{p_i}\right)\Psi,\nonumber \\
  \delta_1I|_2&=&-i\int d^4x\textrm{Tr}\left((\frac{q}{2e^2}
  +\frac{1}{e^2}\frac{\textrm{d}}{\textrm{d}y})(\overline{\epsilon}C_i
  (s_{i1}B_{i1}+
  s_{i2}B_{i2}))\right)\Gamma^{a_i}X_{a_i}\Psi\nonumber\\
  &&-i\int d^4x\textrm{Tr}\left((\frac{q}{2e^2}
  +\frac{1}{e^2}\frac{\textrm{d}}{\textrm{d}y})(\overline{\epsilon}C_i
  (t_{i1}B_{i1}+t_{i2}B_{i2}))\right)\Gamma^{p_i}X_{p_i}\Psi,\nonumber\\
   \delta_0 I^{\prime\prime\prime}&=&\int
   d^4x\frac{i}{e^2}r_{mn}\overline{\epsilon}(\Gamma_m X_n+\Gamma_n
   X_m)\Psi.\label{I3}
\end{eqnarray}
  Using the identities (\ref{identity1}),(\ref{si}) and (\ref{ti})
  and requiring the second order variation of action under susy transformation
  vanish, we can determine the parameter $r_{mn}$ and have
  the following form of $I^{\prime\prime\prime}$
\begin{eqnarray}
  I^{\prime\prime\prime}&=&\frac{1}{2e^2}\int d^4x\left(2\psi'^2
   +(2\psi^{\prime}\tan\psi)'\right)
   \textrm{Tr}\left((C_0+C_1)X_4^2+(C_0+C_2)X_5^2\right.\nonumber\\
   &&\left.+(C_0+C_3)X_6^2+(C_2+C_3)X_7^2+(C_1+C_3)X_8^2+(C_1+C_2)X_9^2\right)\nonumber\\
   &&+\left(2\psi'^2-(2\psi^{\prime}\cot\psi)'\right)
   \textrm{Tr}\left((C_2+C_3)X_4^2+(C_1+C_3)X_5^2\right.\nonumber\\
   &&\left.+(C_1+C_2)X_6^2+(C_0+C_1)X_7^2+(C_0+C_2)X_8^2+(C_0+C_3)X_9^2\right)\nonumber\\
   &&-2(\frac{\psi{\prime}^2}{\cos^2\psi}+\frac{\psi{\prime}^2}{\sin^2\psi})\textrm{Tr}(
   (C_0+C_1)(C_2+C_3)(X_4^2+X_7^2)\nonumber\\
   &&+(C_0+C_2)(C_1+C_3)(X_5^2+X_8^2)
   +(C_0+C_3)(C_1+C_2)(X_6^2+X_9^2)).
\end{eqnarray}
The detailed calculation is given in the appendix.

Similar to the case studied in \cite{HEG0613}\cite{KKL}, there is
enhanced global symmetry $SU(3)$ with $1/4$ supersymmetry when
$C_0=C_1=C_2=C_3=1/4$. For $C_0=C_1=1/2,\ C_2=C_3=0$ the half
supersymmetric configuration has enhanced global symmetry
$SO(2)\times SO(2)$.

   With the same method of obtaining eight supercharges vacuum
structure, we can analyze the vacuum structure of half
supersymmetric configuration.  Without making confusion, we take the
following notation:
\begin{eqnarray}
  \tilde{X}_{i+3}=X_{i+3}(\cos\psi)^{C_0+C_i}(\sin\psi)^{1-C_0-C_i},\
  \ \tilde{F}_{3\ i+3}=\frac{D_3\tilde{X}_{i+3}}
  {(\cos\psi)^{C_0+C_i}(\sin\psi)^{1-C_0-C_i}},\\
  \tilde{X}_{i+6}=X_{i+6}(\sin\psi)^{C_0+C_i}(\cos\psi)^{1-C_0-C_i},\
  \ \tilde{F}_{3\ i+6}=\frac{D_3\tilde{X}_{i+6}}
  {(\sin\psi)^{C_0+C_i}(\cos\psi)^{1-C_0-C_i}},
\end{eqnarray}
  where $i=1,2,3$. With these notations one can simplify the
  expression of the part action $I+I'''$.    Replacing the terms
  $\frac{1}{2}F_{3a}F^{3a}+\frac{1}{2}F_{3p}F^{3p}$ in $I$
  with $\frac{1}{2}\tilde F_{3a}\tilde F^{3a}+\frac{1}{2}\tilde F_{3p}\tilde
  F^{3p}$ and other terms not changing, the modified action of
  $I$ is then identical to the original action $I+I'''$ except for additional
  boundary terms  proportional to $\psi'$ which vanishes if $\psi'=0$
  at infinite.

  Let us consider the case $C_2=C_3=0,C_0+C_1=1$. Taking the ansatz
  $X_5=X_6=0,A_{\mu}=0,A_3=0$, $\Gamma_{3789}=\Gamma_{3567}\epsilon_0=\epsilon_0$
  and the scalars only depending on $x_3$, we obtain the following
  equations,
\begin{eqnarray}
  \sin\psi\tilde{F}_{34}+\cos\psi\tilde{F}_{37}-F_{89}=0,\ & \
  \sin\psi\tilde{F}_{38}+F_{49}=0 \label{vac1}\\
  \sin\psi\tilde{F}_{39}-F_{48}=0,\ & \
  \sin\psi\tilde{F}_{37}-\cos\psi\tilde{F}_{34}=0 \label{vac2}\\
  \cos\psi\tilde{F}_{38}+F_{79}=0,\ & \
  \cos\psi\tilde{F}_{39}-F_{78}=0\label{vac3}
\end{eqnarray}
   Also we
  can obtain these equations from the energy functional
\begin{eqnarray}
  H&=&-\int d^3x\frac{1}{e^2}\textrm{Tr}((\sin\psi \tilde{F}_{37}-\cos\psi\tilde F_{34})^2
    +(\sin\psi \tilde{F}_{38}+F_{49})^2+(\sin\psi
    \tilde{F}_{39}-F_{48})^2\nonumber\\
    &&+(\sin\psi \tilde{F}_{34}+\cos\psi\tilde F_{37}-F_{89})^2+(\cos\psi
    \tilde{F}_{38}-F_{97})^2
    +(\cos\psi \tilde{F}_{39}-F_{78})^2\nonumber\\
    &&+\left(\frac{2}{e^2}\sin\psi\textrm{Tr}(X_4[X_8,X_9])
    +\frac{2}{e^2}\cos\psi\textrm{Tr}(Y_7[Y_8,Y_9])\right)',
\end{eqnarray}
   where we have omitted a boundary term proportional to $\psi'$   which is
   vanishing at infinity after integration. The energy is bounded below by the
   boundary term. When the boundary term vanishes, the minima of the energy
   gives the  equations (\ref{vac1}-\ref{vac3}). The trivial vacuum  is to let all $X$'s
    commute with each other such that $\tilde{X_4},\tilde{X_p}$ are
    just constant. This is similar to the vacuum of generalized Janus configuration.
    However, for the less supersymmetric case, $X_p$ have different dependence on $\psi$ for
different values of $C_0,C_1$. This means that the vacuum is
different for different less susy Janus configurations.

 To get the nontrivial solutions of (\ref{vac1}-\ref{vac3}) seems difficult.
  Since we do not know the dependence of $\psi$ on $y$, we can not
  solve the equation directly. However they look like Nahm equations. From the second equation in (\ref{vac2}),
  we have $\tilde X_4=\tilde
  X_7+c$ with $c$ being a constant. The above equations are simplified as
\begin{eqnarray}
  D_3\tilde X_7=(\tan\psi)^{C_1-C_0}[\tilde X_8, \tilde X_9],\ \
  \sin\psi\cos\psi D_3\tilde X_8=[\tilde X_9, \tilde X_7],\ \
  \sin\psi\cos\psi D_3\tilde X_9=[\tilde X_7, \tilde X_8]\nonumber
\end{eqnarray}
%First there is a trivial solution which $\tilde X_4,\tilde
%X_7,\tilde X_8,\tilde X_9$ are constant matrixes and commute with
%each other. However   So using $\tilde X_p$
%instead of $X_p$ is more nature.
Now let  us prove there is no nontrivial vacuum. Without losing
generality, we can assume the gauge group to be $SU(2)$ and make the
following ansatz,
 \begin{eqnarray}
 \tilde X_7=-i\sigma_1 g_1, \hspace{3ex}\tilde X_8=-ig_2\sigma_2,
 \hspace{3ex}\tilde
 X_9=-ig_3\sigma_3 \\
 g_1=\frac{f_1(y)}{\sin2\psi},\ \ g_2=\frac{f_2(y)}{2\sin\psi^{C_0}\cos\psi^{C_1}},
 \ \ g_3=\frac{f_3(y)}{2\sin\psi^{C_0}\cos\psi^{C_1}}\label{X9}
 \end{eqnarray}
 with $\sigma_i$'s being Pauli matrices.  If $\psi$ is a constant,
 the above set of equations
could be reduced to
\begin{equation}
 f_2f_3=\partial_yf_1,
 \hspace{3ex}f_1f_3=\partial_yf_2,
 \hspace{3ex}f_1f_2=\partial_yf_3,
 \end{equation}
whose solutions are
\begin{eqnarray}
  f_1(y;k,F,y_0)&=&-\frac{Fcn_k(F(y-y_0))}{sn_k(F(y-y_0))} \nonumber
  \\
 f_2(y;k,F,y_0)&=&-\frac{Fdn_k(F(y-y_0))}{sn_k(F(y-y_0))} \nonumber
  \\
 f_3(y;k,F,y_0)&=&-\frac{F}{sn_k(F(y-y_0))}
 \end{eqnarray}
where $sn_k,cn_k,dn_k$ are Jacobi elliptic functions with $k$ being
elliptic modulus and $F\geq 0, y_0$ are arbitrary constants. The
permutations of $f_1,f_2$ and $f_3$ are still the solutions of the
above equations.

However when generically $\psi$ is not a constant and depends on
$y$, it is not easy to solve the equations. Nevertheless let us
start from simple case in which $\psi$ is a constant in the section
($y_j,y_{j+1}$), where $j=1,2,...n$ and $y_j<y_{j+1}$. Namely the
$y_j$'s divide the $x_3$ coordinate into $n+2$ sections and
$\psi(y)$ is a ladder function. In different section, $\psi,F,k$ and
$y_0$ can have different values. Note that $sn_k$ is a periodic
function with period 4K(k), where K(k) is the complete elliptic
integral of the first kind. The function K(k) goes to infinity at
$k=1$. As the zeros of $sn_k(y)$ are $y=0, 2K(k)$, the above
solutions blow up at the zeros of the function $sn_k(F(y-y_0))$. In
order to avoid the infinity in each section, one has to carefully
choose the parameters such that in each section $sn_k(F(y-y_0))$ is
always positive or negative. Recall that $\frac{1}{e^2}=D\sin2\psi$,
so we have $0<\psi<\pi/2,\ \sin\psi,\ \cos\psi>0$.  Since $\tilde
X_7,\tilde X_8,\tilde X_9$ are continuous functions, then $g_2,g_3$
are always positive  or always negative by using (\ref{X9}).
However, this requirement can not be satisfied. In the sections
$y<y_1$ and $y>y_{n+1}$ we must take $k=1$ in order to avoid the
infinity. Then the solution becomes
\begin{eqnarray}
  f_3(y;k=1,F,y_0)&=&-\frac{F\cosh(F(y-y_0))}{\sinh(F(y-y_0))}\\
  f_1(y;k=1,F,y_0)&=&f_2(y;k=1,F,y_0)=-\frac{F}{\sinh(F(y-y_0))},
\end{eqnarray}
so we have $g_2(y_1)>0,g_3(y_1)>0$ and
$g_2(y_{n+1})<0,g_3(y_{n+1})<0$ which are contrast to the above
requirement. This indicates that  we can not find nontrivial
solutions in the case that $\psi$ is a generalic ladder functions.
Since the ladder functions can approach to a general continuous
functions, we can conclude that there is no nontrivial vacuum for
general profile of $\psi(y)$ satisfying \ $0<\psi<\pi/2$. In the
Janus configuration without theta angle, there is no nontrivial
vacuum if there is no point where $e^2$ vanishes \cite{KKL}. In the
point where $e^2$ vanishes, the rescaling scalar need not to be
continuous so that there can exist nontrivial solution. However we
do not have such special point that the rescaling scalars can not be
continuous in the Janus configuration with the theta angle. So
finally we do not have nontrivial vacuum.

Finally, we give the equations for the vacuum preserving two
supercharge, corresponding to the case with $C_0=C_1=C_2=C_3=1/4$,
\begin{eqnarray}
  \sin\psi\tilde F_{34}-F_{56}-F_{89}+\cos\psi\tilde
        F_{37}=0,\nonumber\\
  \sin\psi\tilde F_{35}-F_{64}-F_{97}+\cos\psi\tilde
        F_{38}=0,\nonumber\\
  \sin\psi\tilde F_{36}-F_{45}-F_{78}-\cos\psi\tilde
        F_{39}=0,\nonumber\\
  \sin\psi\tilde F_{37}-F_{59}-F_{68}-\cos\psi\tilde
       F_{34}=0,\nonumber\\
  \sin\psi\tilde F_{38}+F_{49}+F_{67}-\cos\psi\tilde
       F_{35}=0,\nonumber\\
  \sin\psi\tilde F_{39}-F_{48}+F_{57}+\cos\psi\tilde
       F_{36}=0.\nonumber
\end{eqnarray}

\section{Conclusion}

   In this paper we studied several aspects of Janus configurations with $\theta$-angle.
 We discussed the vacuum structure of the original field theory proposed
 in \cite{GW1}, both from the supersymmetry analysis and energy
 functional. It turned out that the vacuum structure is quite
 different from the one of Janus configurations studied in
 \cite{KKL}, where a nontrivial vacuum structure had been discovered.
 We also investigated the BPS solutions of generalized Janus
 configurations. These BPS solutions turns out to be the dyons in
 the field theory, with a nice brane configuration as
 $(p,q)$-strings ending on D3-branes. Finally, we discussed the less
 supersymmetric Janus configurations with $\theta$-angle and proved that it had
 no nontrivial vacuum. We started from the most general projection conditions
 and obtained the Janus configurations with two supercharges. We
 found that in  special cases the  global symmetry got enhanced and the configurations
 had more supersymmetries.

 We tried to solve the half BPS soliton solutions in the Abelian limit in the sharp
 interface case. It would be important to find the solutions
 without taking the abelian limit. And it is also interesting to
 find 1/4 BPS string-junction solutions. For the case with more
 interfaces, the construction of the solution is more complicated.

In the study of the brane configurations corresponding to the BPS
solutions of generalized Janus configuration, we studied the
compatible ways to introduce $(p,q)$-string in (D3, 5)-brane system.
One situation we did not discuss is that when (p,q)-string lie in
the worldvolume of D3-brane, in which case the (p,q)-string and
D3-brane would form bound state\cite{KNP}. This would result in a
noncommutative Janus configuration. It is interesting to construct
such a field theory\cite{HH}.

In \cite{GR,HEG0724,HEG0722,Lunin06}, the half-BPS Janus
supergravity solutions were studied systematically. In this case,
the global symmetry $OSP(4|4)$ is essential to making ansatz to
solve the supergravity equation. For the less supersymmetric case,
the global symmetry is further broken due to the presence of other
5-branes. It would be interesting to construct the corresponding
supergravity solution\cite{Lunin08}.

\section*{Acknowledgments}

The work was partially supported by NSFC Grant
No.10535060,10775002 and NKBRPC (No. 2006CB805905). BC would like
to thank the organizers of Monsoon workshop for the hospitality
and setting up a stimulating discussion section. We would like to
thank Junbao Wu for his participation in the initial stage of the
project.

%\newpage
\appendix
\section{Appendix}

  The second order variation of action can be simplified as
  following by using (\ref{si},\ref{ti})
\begin{eqnarray}
  \delta_1I^{\prime}&=&-i\int d^4x\frac{1}{e^2}\textrm{Tr}\overline{\epsilon}
  (C_i2\psi^{\prime}\tan\psi\Gamma^{a_i}X_{a_i}
  -C_i2\psi^{\prime}\cot\psi\Gamma^{p_i}X_{p_i})\nonumber\\
  &&\left(\frac{1}{2}\psi'B_0+C_jh_j\frac{\psi'}{\sin2\psi}
  (-\sin\psi B_{j1}+\cos\psi B_{j2})\right)\Psi\nonumber\\
  &=&-i\int d^4x\frac{1}{e^2}\textrm{Tr}\left( \overline{\epsilon}
  (-C_i\psi'^2\tan\psi B_0\Gamma^{a_i}X_{a_i}
  +C_i\psi'^2\cot\psi
  B_0\Gamma^{p_i}X_{p_i})\right.\nonumber\\
  &&\left.+C_iC_jh_j\frac{\psi'^2}{\cos^2\psi}\epsilon\Gamma^{a_i}X_{a_i}
  (-\sin\psi B_{j1}+\cos\psi B_{j2})\right.\nonumber\\
  &&\left.-C_iC_jh_j
  \frac{\psi'^2}{\sin^2\psi}\epsilon\Gamma^{p_i}X_{p_i}
  (-\sin\psi B_{j1}+\cos\psi B_{j2})\right)\Psi\label{second1}
\end{eqnarray}
and
\begin{eqnarray}
  \delta_1I|_2&=&-i\int d^4x\textrm{Tr}\left((\frac{q}{2e^2}
  +\frac{1}{e^2}\frac{\textrm{d}}{\textrm{d}y})(\overline{\epsilon}C_i
  2\psi'\tan\psi)\right)\Gamma^{a_i}X_{a_i}\Psi\nonumber\\
  &&+i\int d^4x\textrm{Tr}\left((\frac{q}{2e^2}
  +\frac{1}{e^2}\frac{\textrm{d}}{\textrm{d}y})(\overline{\epsilon}C_i
  2\psi'\cot\psi)\right)\Gamma^{p_i}X_{p_i}\Psi\nonumber\\
  &=&-\frac{i}{e^2}\int d^4x
  C_i\left(q\psi'\tan\psi+(2\psi'\tan\psi)'\right)
  \overline{\epsilon}\Gamma^{a_i}X_{a_i}\Psi\nonumber\\
  &&+C_i(\psi'^2\tan\psi)
  \overline{\epsilon}B_0\Gamma^{a_i}X_{a_i}\Psi\nonumber\\
  &&+C_i\left(-q\psi'\cot\psi-(2\psi'\cot\psi)'\right)
  \overline{\epsilon}\Gamma^{p_i}X_{p_i}\Psi\nonumber\\
  &&+C_i(-\psi'^2\cot\psi)
  \overline{\epsilon}B_0\Gamma^{p_i}X_{p_i}.\Psi\label{second2}
\end{eqnarray}
Note that $\overline{\epsilon}\Gamma^{a_i}X_{a_i} (-\sin\psi
B_{j1}+\cos\psi B_{j2})$
  are proportion to
  $\overline{\epsilon}(\sin\psi B_{j1}+\cos\psi B_{j2})\Gamma^{a_i}X_{a_i}$,
  using the project condition (\ref{identity1}), they are proportion
  to $\overline{\epsilon}h_j\Gamma^{a_i}X_{a_i}$ which is the same form of (\ref{I3}).
  Since susy require $\delta I^{\prime\prime\prime}+\delta_1 I'+\delta_1
  I|_2=0$, such terms should cancel each other. However, we have
  another form $\overline{\epsilon}B_0\Gamma^{a_i} X_{a_i}$. These
  terms cancel each other in (\ref{second1})(\ref{second2}).

\end{document}